\documentclass[journal]{IEEEtran}
\usepackage{graphicx}

\newcommand{\Aprimebold}{\ensuremath{\mathrm{\mathbf{A}}^\prime}}
\newcommand{\Aprime}{\ensuremath{\mathrm{A}^\prime}}
\newcommand{\fluenceunit}{1~MeV~neutron~equivalent/cm\ensuremath{^2}}

\begin{document}
\title{The Silicon Vertex Tracker for the \\Heavy Photon Search Experiment}

\author{Per~Hansson~Adrian$^1$, on behalf of the HPS Collaboration\\
$^1$SLAC National Accelerator Laboratory, Menlo Park, CA, USA

\thanks{Manuscript received November 23, 2015. Work supported by 
the U.S. Department of Energy under contract number DE-AC02-76SF00515, 
the National Science Foundation,  
French Centre National de la Recherche Scientifique and 
Italian Istituto Nazionale di Fisica Nucleare. Authored by Jefferson Science 
Associates, LLC under under U.S. Department of Energy contract No. DE-AC05-06OR23177.}%
%\thanks{Rouven Essig is supported in part by the Department of Energy Early Career research program 
%DESC0008061and by a Sloan Foundation Research Fellowship. Authored by Jefferson Science 
%Associates, LLC under under U.S. Department of Energy contract No. DE-AC05-06OR23177.
}

\maketitle
\pagestyle{empty}
\thispagestyle{empty}

\begin{abstract}
The Heavy Photon Search (HPS) is a new, dedicated experiment at Thomas Jefferson National Accelerator Facility (JLab) to search for a massive vector boson, the heavy photon (a.k.a. dark photon, \Aprimebold{}), in the mass range 20-500~MeV/c$^{2}$ and with a weak coupling to ordinary matter. An \Aprimebold{} can be radiated from an incoming electron as it interacts with a charged nucleus in the target, accessing a large open parameter space where the \Aprimebold{} is relatively long-lived, leading to displaced vertices. HPS searches for these displaced \Aprimebold{} to e$^+$e$^-$ decays using actively cooled silicon microstrip sensors with fast readout electronics placed immediately downstream of the target and inside a dipole magnet to instrument a large acceptance with a relatively small detector. With typical particle momenta of 0.5-2~GeV/c, the low material budget of 0.7\% $\mathbf{X_0}$ per tracking layer is key to limiting the dominant multiple scattering uncertainty and allowing efficient separation of the decay vertex from the prompt QED trident background processes. Achieving the desired low-mass acceptance requires placing the edge of the silicon only 0.5~mm from the electron beam. This results in localized hit rates above 4~MHz/mm$^2$ and radiation levels above $\mathbf{10^{14}}$ 1~MeV neutron equivalent /cm$^2$ dose. Hit timing at the ns level is crucial to reject out-of time hits not associated with the \Aprimebold{} decay products from the almost continuous CEBAF accelerator beam. To avoid excessive beam-gas interactions the tracking detector is placed inside the accelerator beam vacuum envelope and is retractable to allow safe operation in case of beam motion. This contribution will discuss the design, construction 
and first performance results from the first data-taking period in the spring of 2015. 
\end{abstract}

%\begin{IEEEkeywords}
%IEEEtran, journal, \LaTeX, paper, template.
%\end{IEEEkeywords}

\section{Introduction}

\IEEEPARstart{R}{ecent} astrophysical results ~\cite{pamela,fermi} have generated intense interest in physics models 
beyond the Standard Model with a new force, mediated by a massive, sub-GeV scale, 
U(1) gauge boson (a.k.a. the Heavy Photon, Dark Photon or \Aprime{}) that couples very weakly to 
ordinary matter through "kinetic mixing"~\cite{nima,holdom}. The existence of such a new force is in 
accord with astrophysical and cosmological constraints. Its weak coupling to the electric charge could be 
the only non-gravitational window into the existence of hidden sectors consisting of particles that do not 
couple to any of the known forces~\cite{Hewett:2012ns}.

\section{The Heavy Photon Search Experiment}
\begin{figure}[]
\centering
\includegraphics[width=3.5in]{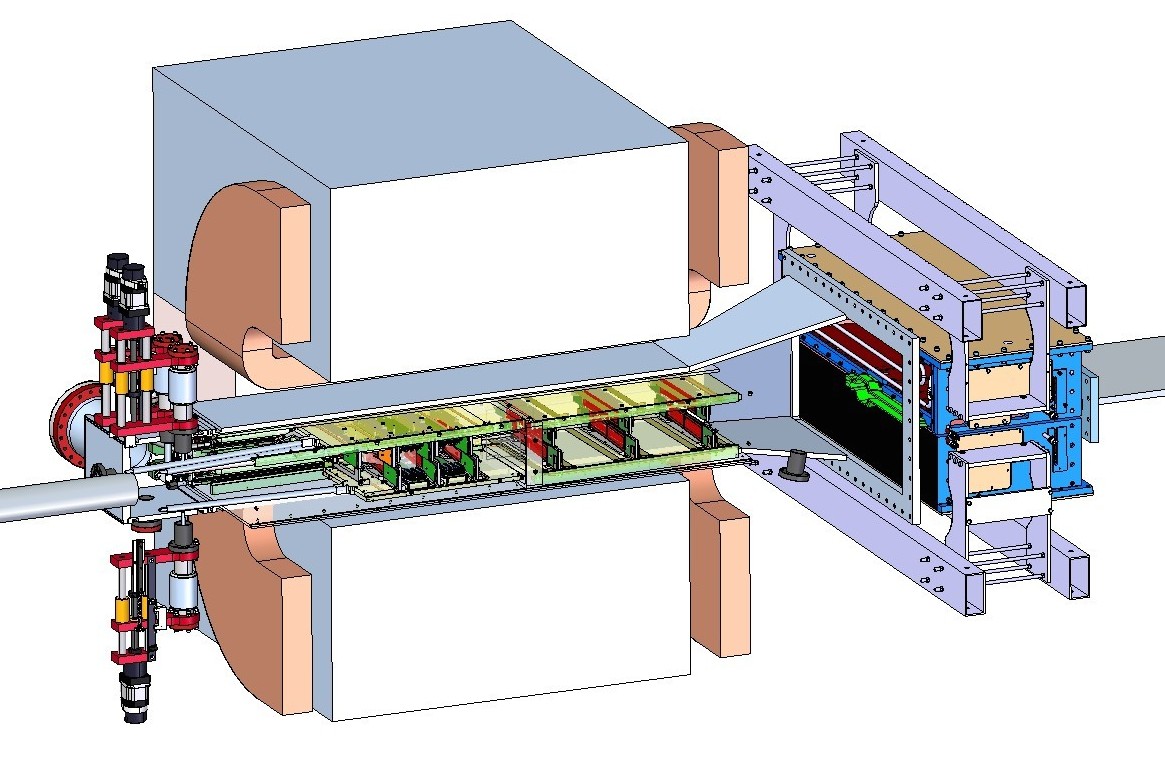}
\caption{A rendered overview of the HPS detector.}
\label{fig:hps-layout}
\end{figure}

The Heavy Photon Search experiment (HPS) is a new fixed-target experiment~\cite{proposal_full}
specifically designed to discover an \Aprime{} with m$_{\Aprime}=20-500$~MeV, produced through bremsstrahlung 
in a tungsten target and decaying into an $e^{+}e^{-}$ pair.  
In particular, the HPS experiment has sensitivity to the challenging region with small cross sections out of 
reach from collider experiments and where thick absorbers, as used in beam-dump experiments to 
reject backgrounds, are ineffective due to the relatively short \Aprime{} decay length ($<1$~m)~\cite{bible}.  
This is accomplished 
by placing a compact silicon tracking and vertex detector (SVT) in a magnetic field, immediately downstream (10~cm) 
of a thin ($\sim 0.125\%~X_{0} $) target to reconstruct the mass and decay vertex position of  the \Aprime{}. A 
rendered overview of HPS is shown in Fig.~\ref{fig:hps-layout}.

HPS runs in Hall~B at Thomas Jefferson National Accelerator Facility (JLab) using the CEBAF 
accelerator electron beam with an energy of 1.05~GeV and 50~nA current, with planned operation of up to 6.6~GeV and 450~nA. 
The kinematics of \Aprime{} production 
typically result in final state particles within a few degrees of the beam, especially at low $m_{\Aprime}$ . 
Because of this, the detector must accommodate passage of the beam 
downstream of the target and operate as close to the beam as possible. Because background rates in this region from the 
scattered beam are very large, a fast lead-tungstate crystal calorimeter trigger with 250~MHz flash ADC readout~\cite{fadc250} and 
excellent time tagging of hits is used to trigger on interesting events and reduce the bandwidth required to transfer data from the 
detector.  This method of background reduction is the motivation for operating HPS in a
nearly continuous beam: in a beam with large per-bunch charge, background from a single bunch would fully occupy the detector 
at the required beam intensity.

\section{The Silicon Vertex Tracking Detector}
The Silicon Vertex Tracker (SVT) allows for precise and efficient reconstruction of charged particles and their trajectories. 
%The design was driven mainly by the specific physics signature of \Aprime{} production and decay and the constraints from the environment 
%at the interaction region. 
At beam energies between 1.0-6.6~GeV, the electron and positron from the \Aprime{}  
decay will be produced with momenta in the range of 0.4-2~GeV$/c$ and angles of 10-100~mrad from the beam. The dominant 
tracking uncertainty in this regime is multiple Coulomb scattering, so the SVT needs to minimize the amount of material in the tracking 
volume. With an approximate goal of 2\% mass resolution in the 1~m long tracking volume (determined by an existing magnet and vacuum 
chamber) with 0.25~T magnetic field (for 1~GeV beam energy); 1\%~$X_0$ or less material per 3D tracking hit and 
six layers was deemed adequate. 
For weak couplings, the \Aprime{} may be long-lived and the $e^{+}e^{-}$ pair decay vertex might be displaced several 
cm downstream of the target foil. To discover rare \Aprime{} displaced decays, the SVT typically needs a prompt rejection of roughly 
10$^7$ at 1~cm vertex resolution~\cite{proposal_full}. In order to reach that performance, the first layer of the SVT needs to be placed 
10~cm from the target. At that distance, the large hit rates from beam electrons undergoing Coulomb scattering in the target allow placing 
the first layer 1.5~mm from the beam. No instrumentation can be placed inside that 15~mrad angle creating a "dead zone" throughout the 
experiment. The expected radiation dose peaks at $10^{15}$~electrons/cm$^2$/month, or roughly 
$3 \times 10^{13}$~\fluenceunit{}/month~\cite{dose}, close to the beam and places further constraints on the sensor 
technology.  Furthermore, the whole tracker has to operate in vacuum to avoid secondary backgrounds from 
beam gas interactions, and have retractable tracking planes and easy access for sensor replacement to increase safety. 
Given the high hit density, the fast time response, and good resolution and radiation hardness needed; silicon microstrip 
sensors are the technology of choice for the tracker. Pixel sensors suitable for instrumenting our large acceptance are either too slow
or have an unacceptable material budget.

\subsection{Layout}
\begin{figure}[]
\centering
\includegraphics[width=3.5in]{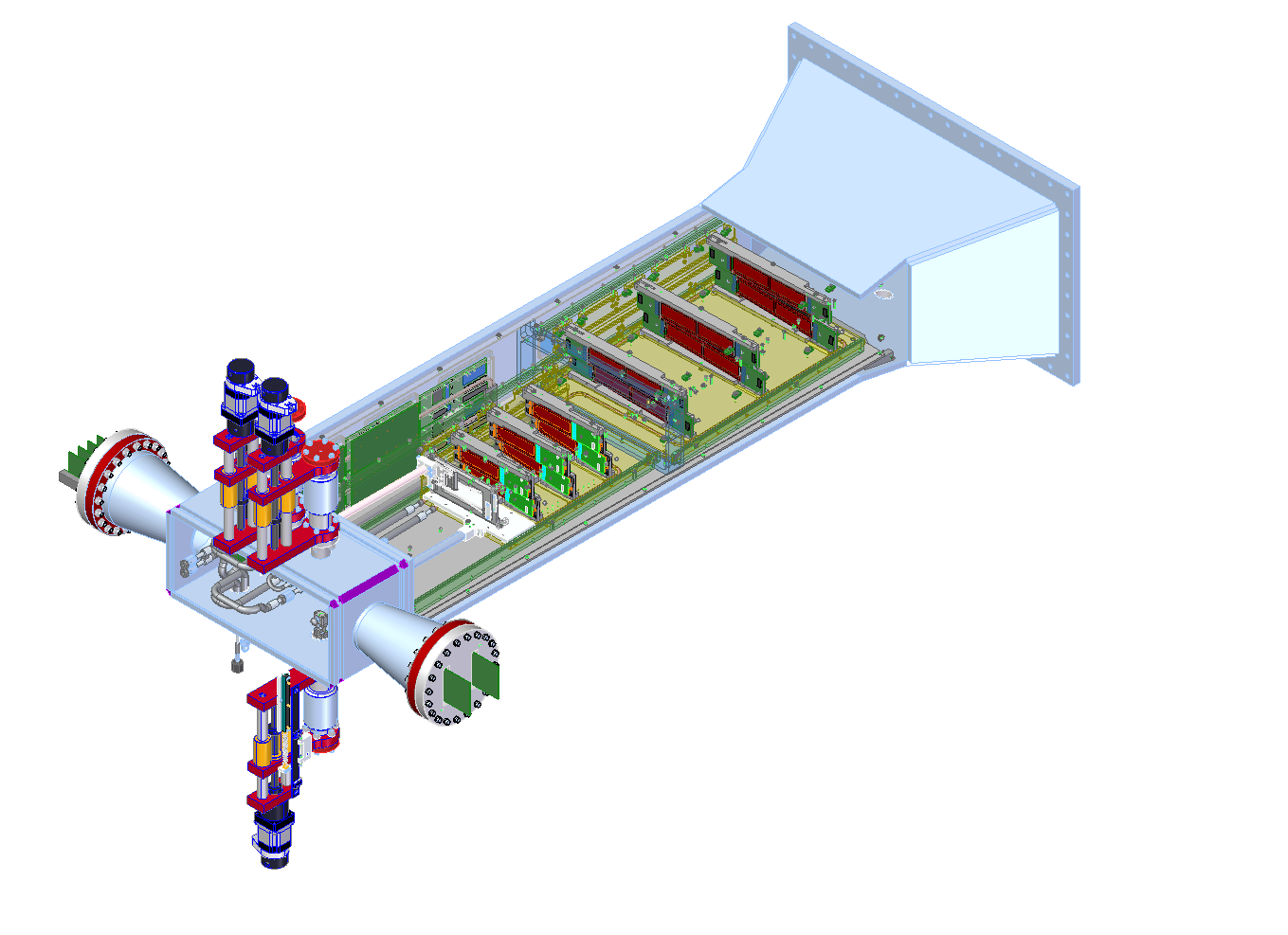}
\caption{A rendered overview of the SVT installed on the beamline.}
\label{fig:layout}
\end{figure}

\begin{figure}[]
\centering
\includegraphics[width=3in]{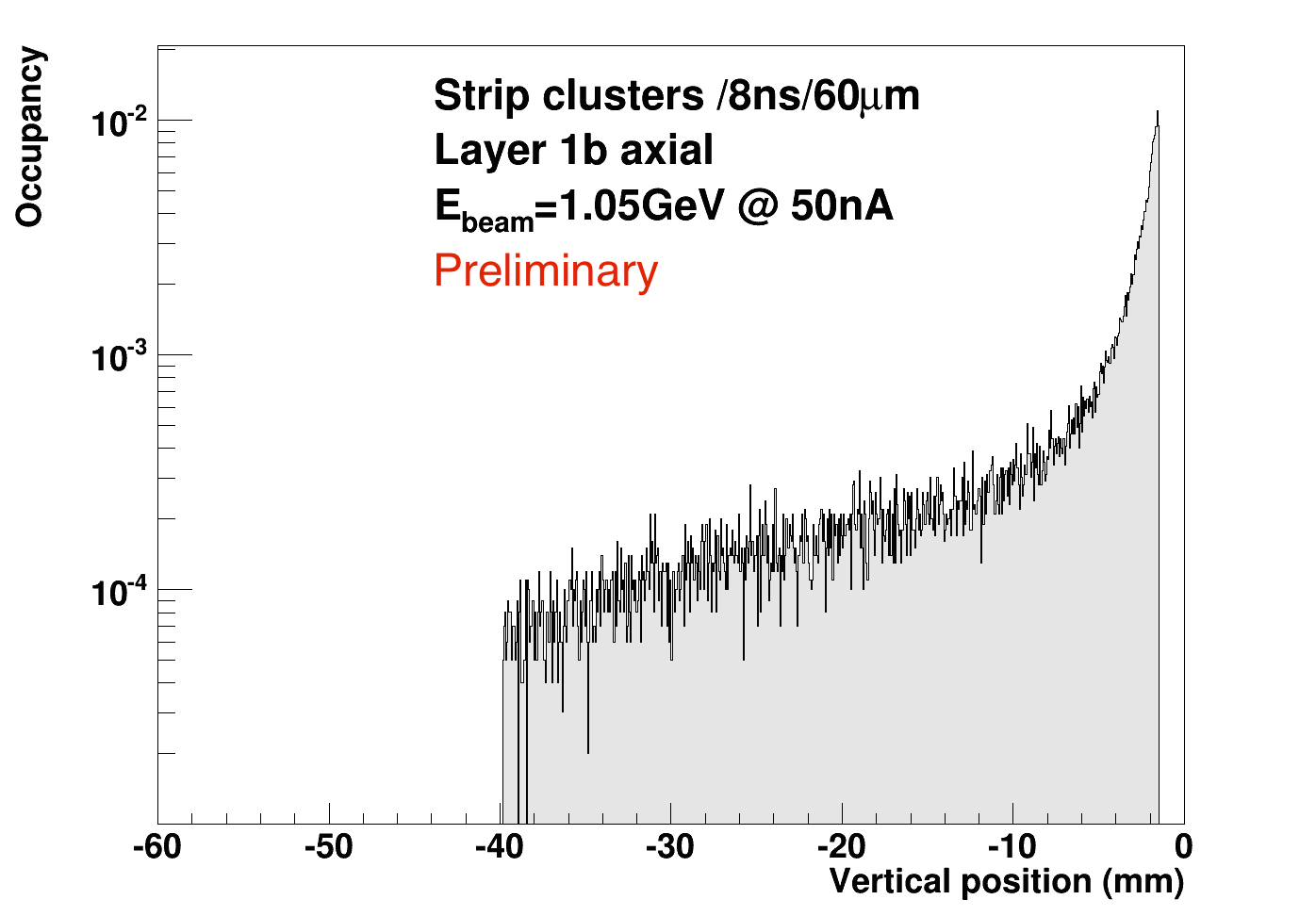}
\caption{Background occupancy for layer 1 during nominal operation conditions.}
\label{fig:occupancy}
\end{figure}

The SVT overall layout is rendered in Fig.~\ref{fig:layout} and summarized in Tab.~\ref{tab:layout}. 
\begin{table}
\centering
\begin{tabular}{|lcc|}
\hline
Layer $\rightarrow$& 1-3 & 4-6 \\ 
\hline
$z$ pos. (cm)  & 10-30 & 50-90  \\
Stereo angle (mrad)  & 100 & 50  \\
Non-bend plane resolution ($\mu$m)  & $\approx 6$ & $\approx6$  \\
Bend plane resolution ($\mu$m)  & $\approx 60$ & $\approx 120$  \\
\hline
\end{tabular}
\caption{Main tracker parameters.}
\label{tab:layout}
\end{table}
Each of the six tracking layers, 
arranged in two halves both above and below the beam to avoid the "dead zone", consists of silicon microstrip sensors placed back-to-back. 
The first three layers have a 100~mrad stereo angle between the sensors and the last three have 50mrad in order to improve the pointing 
resolution to the vertex. The first layer is located only 10~cm downstream of the target to give excellent 3D vertexing performance which, 
with the 15~mrad dead zone above and below the beam axis, puts the active silicon only 1.5~mm from the center of the beam. Hit densities 
in the most active region reach 4~MHz/mm$^2$ and about 1\% occupancy for the strips closest to the beam, see Fig.~\ref{fig:occupancy}.

\subsection{Sensors and Front-End Readout}
The sensors are $p+$-on-$n$, single-sided, AC-coupled, polysilicon-biased microstrip sensors fabricated by Hamamatsu Photonics Corporation 
for the cancelled D\O~Run~2b upgrade~\cite{d0run2b}. These $320~\mu$m thick sensors are $4\times10$~cm$^2$ with 30 and 60~$\mu$m pitch 
for sense and readout strips, respectively,  matching the required material budget and single hit spatial resolution. 
The sensors were qualified to withstand at least 1~kV bias in order to tolerate the $1.5\times10^{14}$~\fluenceunit{} for a 
six month run without significant degradation. 

\begin{figure}[]
\centering
\includegraphics[width=3.0in]{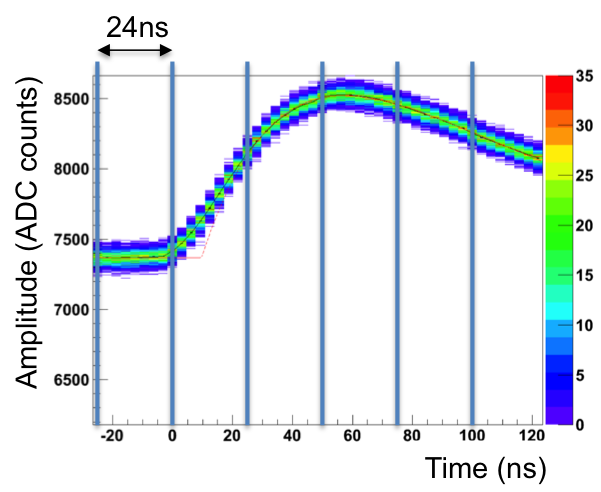}
\caption{Pulse shape of the APV25 ASIC.}
\label{fig:pulseshape}
\end{figure}

One of the key requirements for the SVT is hit time resolution of  $<2$~ns in order to reject background and improve pattern recognition 
accuracy for close to the beam where occupancies are high. This is achieved by using the APV25 front-end readout ASIC~\cite{apv25},  
developed for the 
Compact Muon Solenoid experiment at the Large Hadron Collider. The APV25 is an analog pipeline ASIC with 128 channels of preamplifier 
and shaper, feeding a 192 long analog memory pipeline. In the so-called "multi-peak" readout mode, the APV25 presents 
three consecutive samples of the pulse height in response to an APV25 readout trigger signal. 
By sending two APV25 readout triggers for every event trigger signal 
from the electromagnetic calorimeter, six analog samples of the pulse shape, see Fig.~\ref{fig:pulseshape}, are obtained at a sampling rate of 
41~MSPS. This pulse shape can be analyzed and fitted to extract the t$_0$ of the hit~\cite{Friedl:2009zz}.
The main parameters of the APV25 ASIC are shown in Tab.~\ref{tab:apv25}: the 44~$\mu$m pitch, low noise and operation using either polarity together with the 
proven robustness and radiation hardness is a good fit for HPS. 
\begin{table}
\centering
\begin{tabular}{|l|c|}
\hline
Technology & 0.25~$\mu$m \\ 
\hline
Channels & 128 \\
\hline
Input pitch & 44~$\mu$m \\
\hline
Noise [ENC e$^-$] & $270 + 36\times$C~(pF)\\
\hline
Power consumption & 350mW \\
\hline
\end{tabular}
\caption{Main APV25 ASIC parameters.}
\label{tab:apv25}
\end{table}
The sensor and APV25 chip can be seen in Figure~\ref{fig:half-modules-L1-3}.

\subsection{Module Design}
\begin{figure}[]
\centering
\includegraphics[width=3.0in]{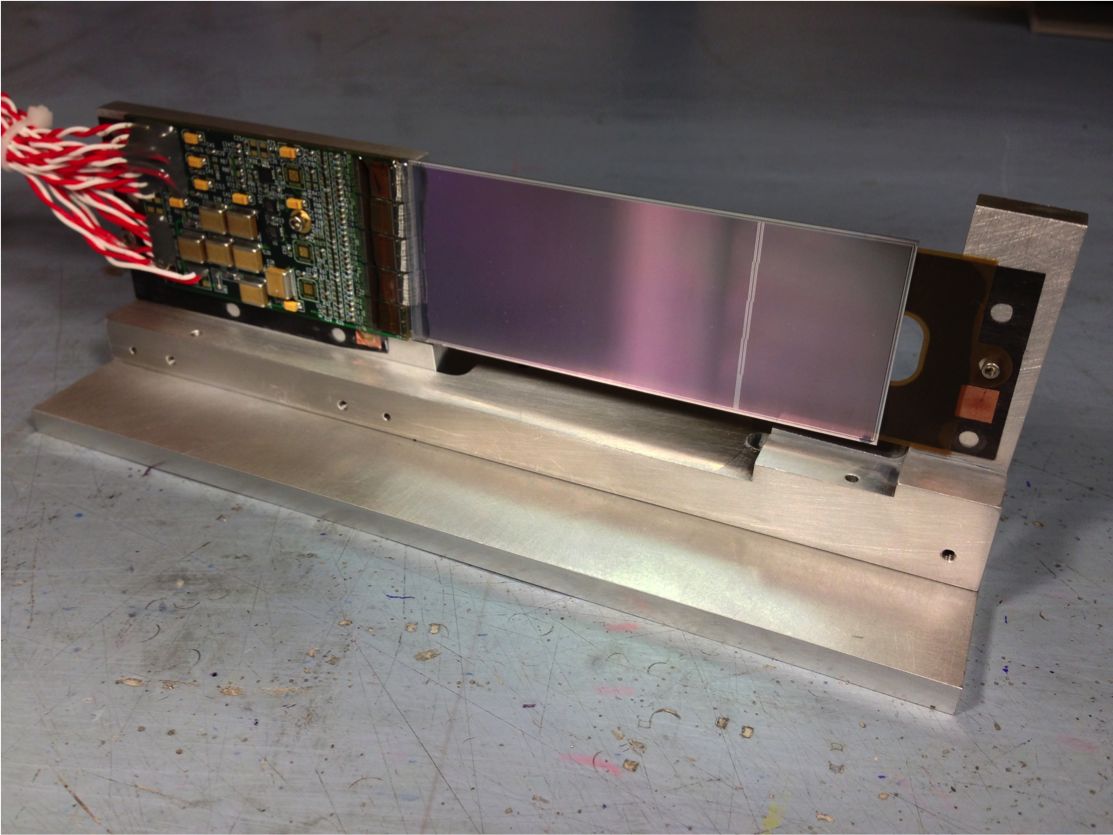}
\caption{A half-module for layer 1-3 mounted on the module support.}
\label{fig:half-modules-L1-3}
\end{figure}

\begin{figure}[]
\centering
\includegraphics[width=3.0in]{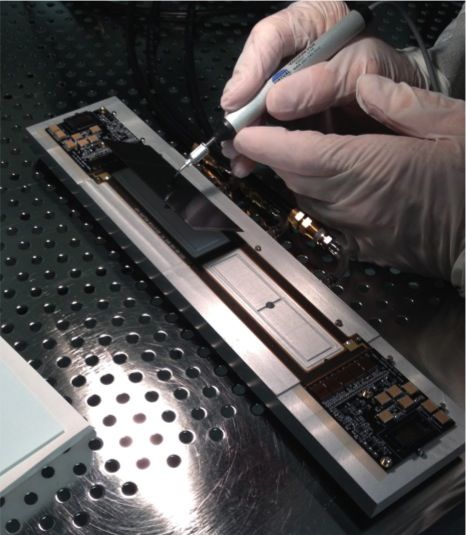}
\caption{A layer 4-6 half-module under assembly.}
\label{fig:half-modules-L4-6}
\end{figure}

Five APV25 chips are mounted on a FR4 hybrid board and wire-bonded to the end of the sensor. 
The hybrid boards, located outside the tracking volume, 
provide power and filtering circuitry for the high voltage bias and a temperature sensor. A hybrid and a sensor, glued to a polyimide-laminated 
carbon fiber composite backing, make up a so-called "half-module" for the first three layers and were re-used from the HPS Test 
detector~\cite{paper_testrun}. 
To increase acceptance, the half-modules for layers 4-6 have two sensors mounted end-to-end and are read out from opposite ends. 
Space constraints required these hybrids to be slightly 
smaller and instead of Teflon-coated twisted pair wires soldered directly to the FR4 board as in layer 1-3, these 
have high density connectors to transfer signal, power and high voltage bias on and off the hybrid. 
Figures~\ref{fig:half-modules-L1-3} and~\ref{fig:half-modules-L4-6} show a complete layer 1-3 and a 4-6 half-module under assembly, 
respectively.

Modules are built by placing two half-modules back-to-back sandwiched around a aluminum cooling block at the hybrid end. The required 
3D space point resolution is obtained by rotating one half-module by 50 or 100~mrad. For layer 1-3 modules, the 
cooling block serves as the fixed part of a support where the opposite end of the half-modules are screwed to an aluminum 
bar with a pivot engaging a screw-adjusted spring, as shown in Fig.~\ref{fig:half-modules-L1-3}. 
This spring-tensioned module support allows holding the sensors, with their carbon fiber backing, straight with very little support material 
and absorbs thermal expansion mismatch. The same principle is used for layer 4-6 where the cooling block at one end has an integrated 
spring pivot.
\begin{figure}[]
\centering
\includegraphics[width=3.0in]{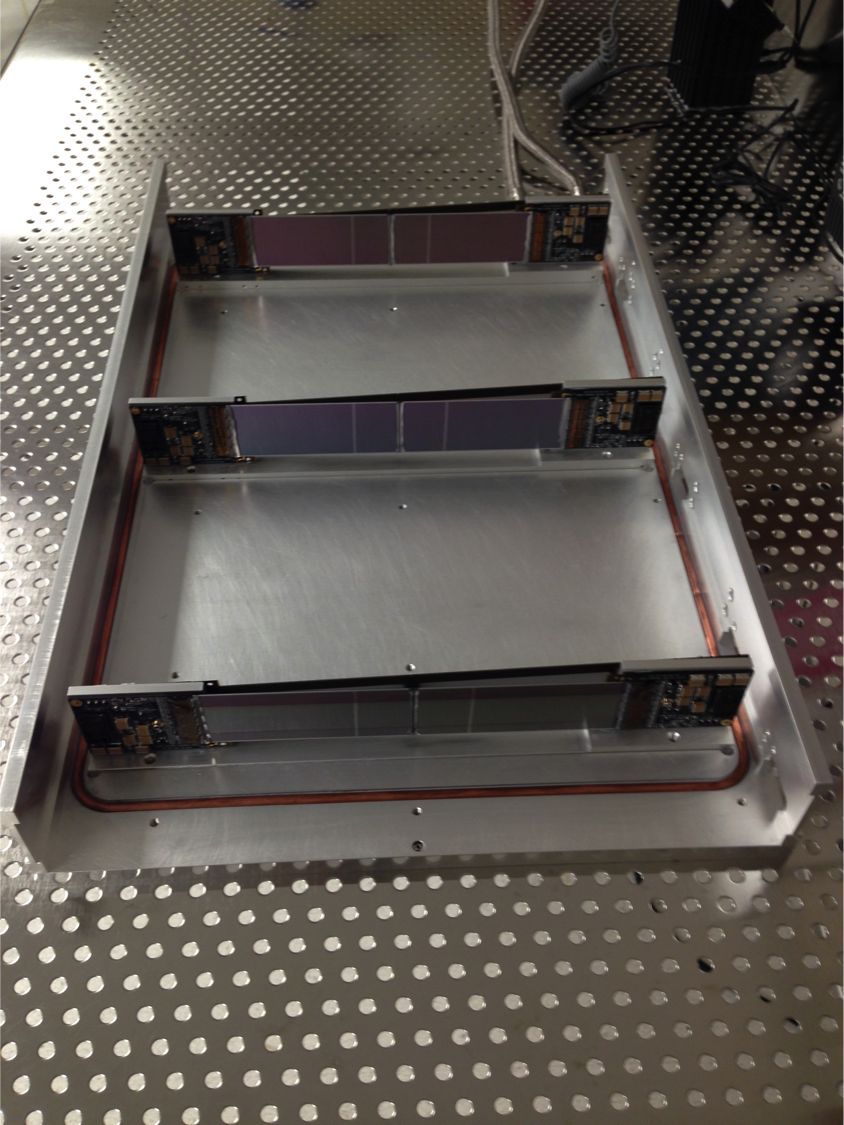}
\caption{A u-channel for layer 4-6 holding three detector modules.}
\label{fig:uchannels-L4-6}
\end{figure}

\subsection{Support and Cooling}
\begin{figure}[!t]
\centering
\includegraphics[width=3.0in]{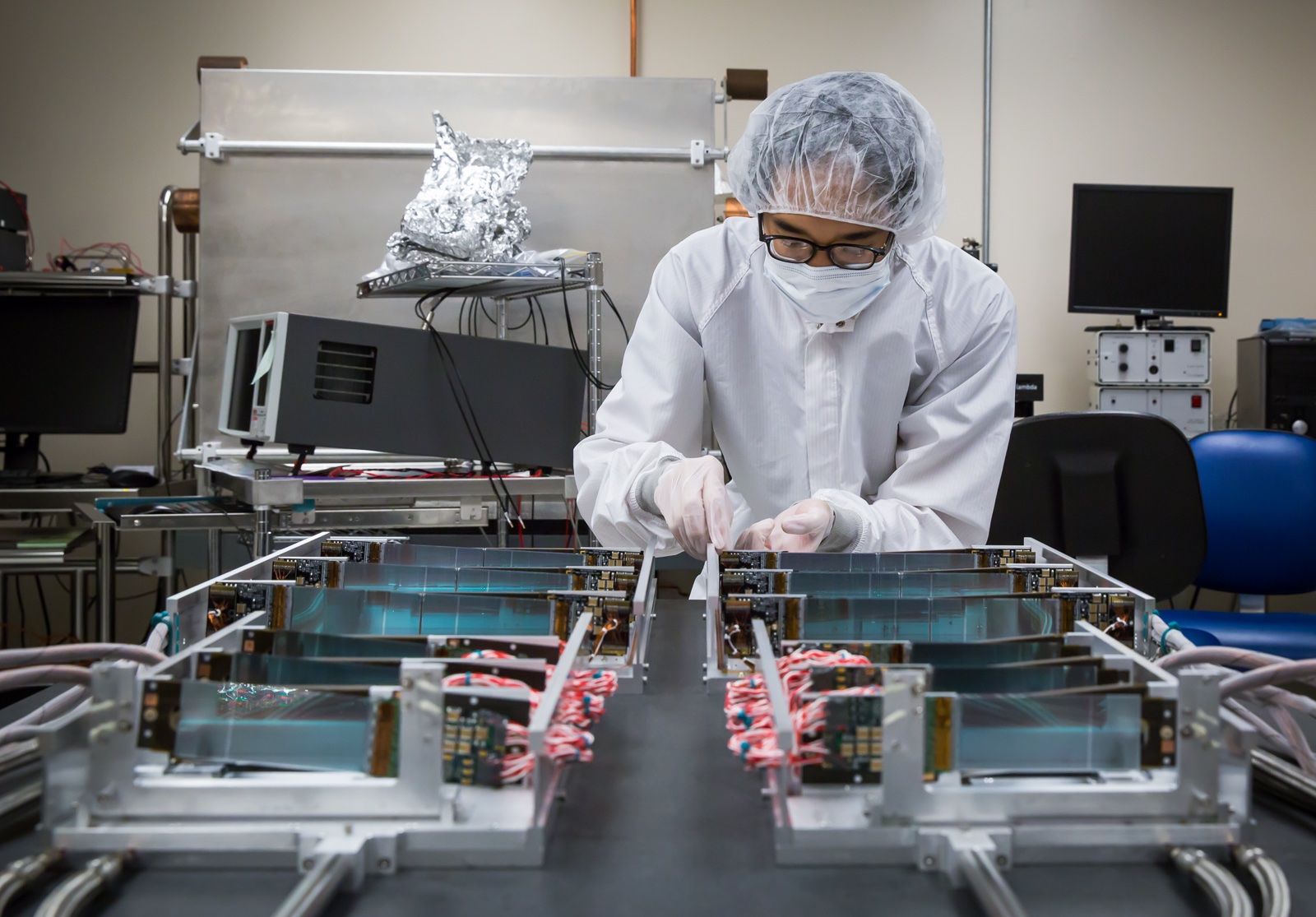}
\caption{All four u-channels (layers 1-3 closest) with detector modules mounted. The flexible cooling hoses and support rods for layer 1-3 
can be seen extending towards the camera view. }
\label{fig:uchannels}
\end{figure}

\begin{figure}[]
\centering
\includegraphics[width=3.0in]{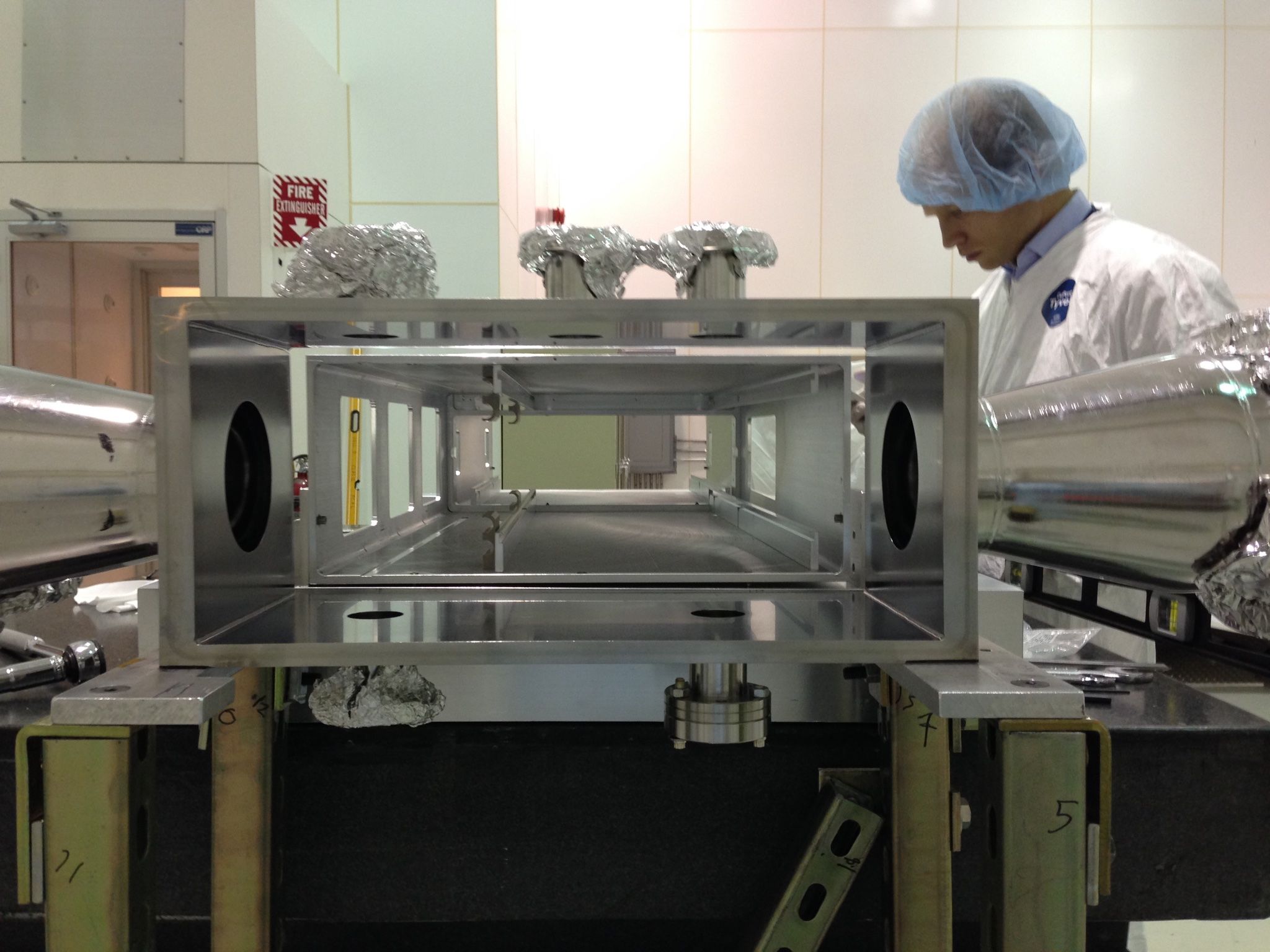}
\caption{The SVT support box with the upstream vacuum box for services.}
\label{fig:svt-box}
\end{figure}

\begin{figure}[]
\centering
\includegraphics[width=3.0in]{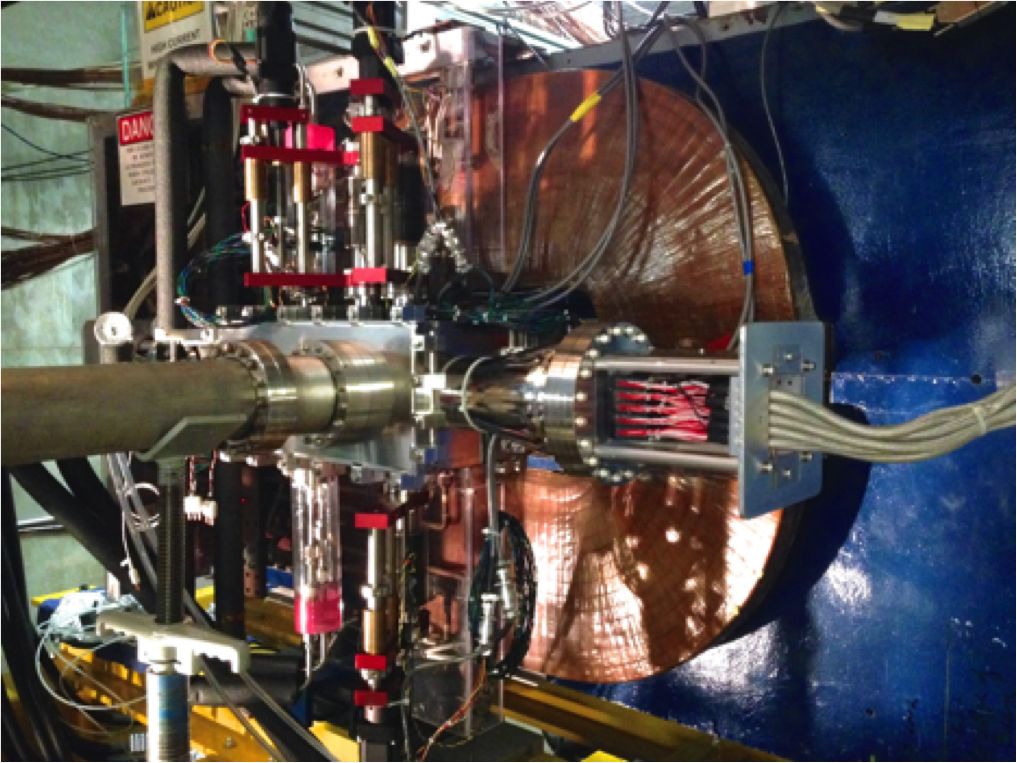}
\caption{View from upstream, electron side, of the SVT after installation on the beamline. The vacuum box, that interfaces the beamline with the 
vacuum chamber, can be seen with its three linear shifts, two on the top and one on the bottom side. The flange holding the power and 
high-voltage sensor bias vacuum penetration boards extends to the right from the support box.}
\label{fig:vacuum-box}
\end{figure}

Three modules are screwed directly on an aluminum "u-channel", see Fig.~\ref{fig:uchannels-L4-6} and~\ref{fig:uchannels}, that are 
actively cooled by HFE~7000 coolant circulating in a 1/4" copper tube pressed into a machined groove. 
Four U-channels, two for layer 4-6 and two for layer 1-3, roll into a rigid support box on guide rails to precision kinematic mounts. The support box, 
shown in Fig.~\ref{fig:svt-box}, sits inside the JLab Hall~B analyzing magnet vacuum chamber on four adjustable supports. In order to provide the 
rigidity needed for precision mounting of the u-channels, a rectangular  "support ring" is located at the inner mounting point for the u-channels. 
The upstream end of the layer 1-3 u-channels is attached to a rigid support 
rod extending upstream to vertical linear shifts. Moving the vertical linear shifts allows the layer 1-3 u-channels to rotate up to 1.3$^{\circ}$ and the 
layer 1 sensors to move more than 7~mm from the beam, allowing clean passage of the beam during beam setup. 
The linear shifts, one for top and one for bottom, penetrate vacuum through bellows and are remote-controlled by a precision stepper motor. The 
sensors closest to the beam can be positioned with 6~$\mu$m steps and $<50~\mu$m reproducibility.
A similar linear shift setup is used to hold and move the target in and out of the beam.
The  flanges for vacuum penetration are located on a custom support box that attaches to the upstream end of the 
magnet as seen in Fig.~\ref{fig:vacuum-box}. In addition to the linear shift flanges, it has flanges for cooling lines and electrical services and the 
upstream interface to the beamline. 

\begin{figure}[]
\centering
\includegraphics[width=3.0in]{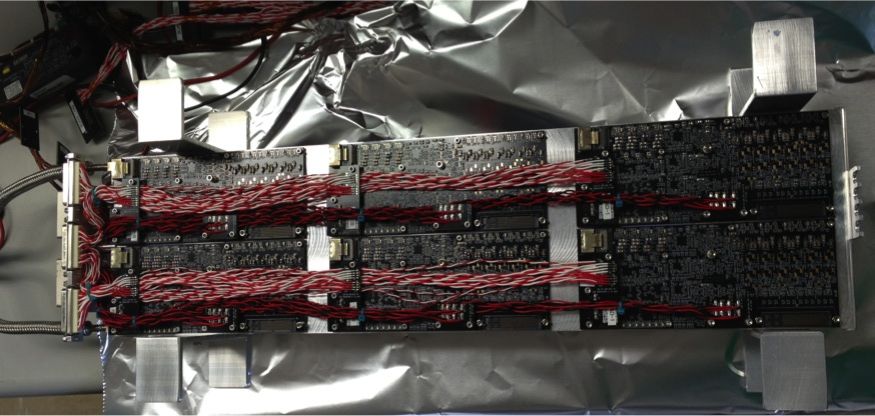}
\caption{The partially cabled data acquisition front end boards  screwed to the aluminum support plate before installation .}
\label{fig:febs}
\end{figure}

An aluminum plate with data acquisition boards, see Fig.~\ref{fig:febs}, slides into the chamber in a machined groove in the support box. 
Coolant is circulated through the u-channels (at about \mbox{-20$^{\circ}$~C}) keeping the silicon at approximately  
\mbox{-10$^{\circ}$~C} to improve performance after radiation damage. Water at 20$^{\circ}$~C is used to cool the data acquisition boards.

\subsection{Data Acquisition}

Analog samples at 41.667~MSPS from the APV25 chips are sent on twisted pair magnet wire 
to a total of 10 Front End Boards (FEB) seen in Fig.~\ref{fig:febs}. Each FEB digitizes and transfers data, from up to four hybrids, 
at up to 3.3~Gb/s using high-speed serial links to Xilinx Zynq based data processing modules on the ATCA based SLAC RCE 
platform~\cite{paper_testrun} for zero suppression and event building.
Each FEB also handles power regulation and monitoring as well as high voltage sensor bias distribution to each of the attached 
hybrids. To shorten the analog signal distance, the FEBs are placed inside the vacuum chamber, pressed against thermal pads on each side of a 1/2$"$ cooled support plate on the 
upstream positron side, rendering a less intense radiation environment. Borated high-density polyethylene is used to 
further lower the risk of damage from radiation emitted by the nearby target.
\begin{figure}[]
\centering
\includegraphics[width=3.0in]{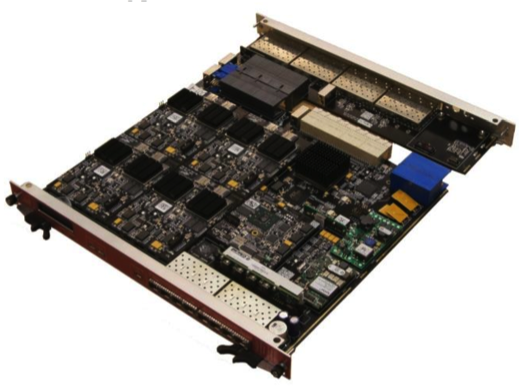}
\caption{One COB ATCA blade used in the RCE platform.}
\label{fig:cob}
\end{figure}
Data from the FEBs are routed via short, flexible miniSAS cables to four flange boards. These are FR4 boards potted through slots in the 
8" vacuum flange on the upstream positron side of the vacuum box. On the out side of the boards, signals are converted to optical and transferred to 
the DAQ platform about 30~m away. A similar mechanical technique is used on the opposite side of the vacuum box to bring in low voltage power 
and high voltage sensor bias into the chamber; this can be seen in Fig.~\ref{fig:vacuum-box}.

\begin{figure}[]
\centering
\includegraphics[width=3.5in]{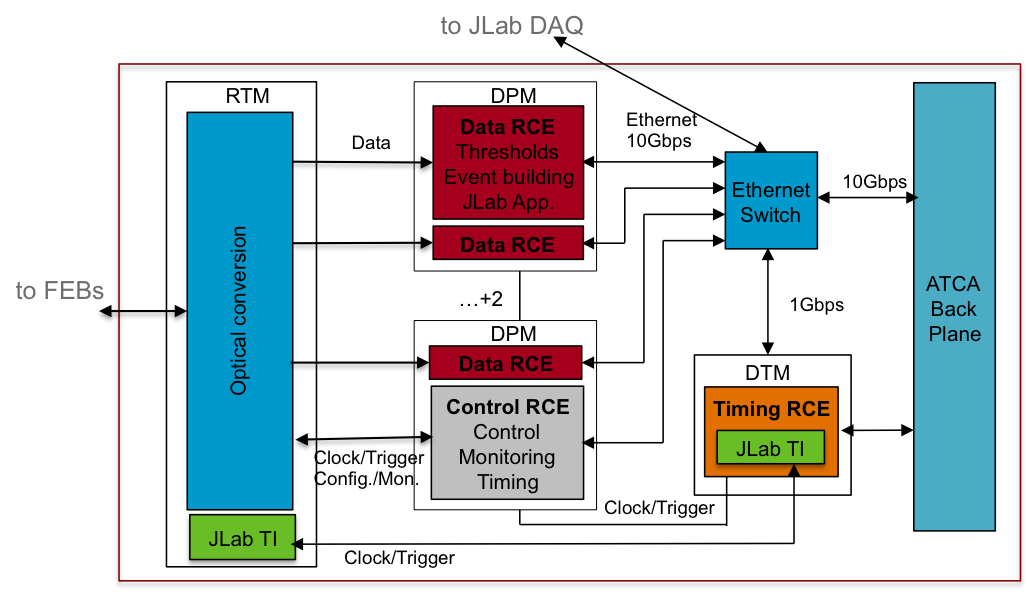}
\caption{Schematic overview of the SVT DAQ.}
\label{fig:rce}
\end{figure}

An overview of the data flow across the RCE platform is shown in Fig.~\ref{fig:rce}.
Data from 10 FEBs are split and sent to 14 processing nodes on two ATCA blades, called COBs (Cluster on board). 
The processing nodes, known as Reconfigurable Cluster Elements (RCE) are based on Xilinx Zync 7000 series 
system-on-chip which has a dual ARM Cortex A9 processor tightly coupled to a 28nm FPGA fabric. The independent operating nodes receive data 
from up to four hybrids, apply calibrated thresholds and build event frames in the firmware. A readout application from the JLab DAQ~\cite{coda}, 
running on the ARM processor, pulls the event frames from memory via DMA and transfers the event frames to the JLab DAQ 
event builder over 10~Gb/s ethernet. 
The COB also hosts a special RCE that handles the trigger and timing distribution across the processing nodes on each COB. 
This RCE implements the JLab trigger interface 
firmware and accepts the master clocks together with trigger information from the JLab DAQ from a special fiber attached on the 
custom rear transition board.  One of the RCEs is allocated to handle control, trigger and timing signals to and from all the hybrids. It also hosts the 
slow control and environmental interfaces to the EPICS control system. 

During running the system operated at about 20~kHz and with data rates up to 150~MB/s. It has been tested to 50~kHz and 200~MB/s.

\subsection{Performance}
The first run of HPS with an electron beam was carried out in the spring of 2015. A primary goal of this engineering run was to prove the 
operational principles. Thanks to a 2-week extension of the running schedule HPS was able to take data with beam current and beam size as 
desired and the SVT in its nominal position with sensor edges only 0.5~mm from 
the the center of the beam.  In all, roughly 1/3 of a week of physics data was taken at the nominal operating point with 1.05~GeV 
beam energy. 

\begin{figure}[]
\centering
\includegraphics[width=3.0in]{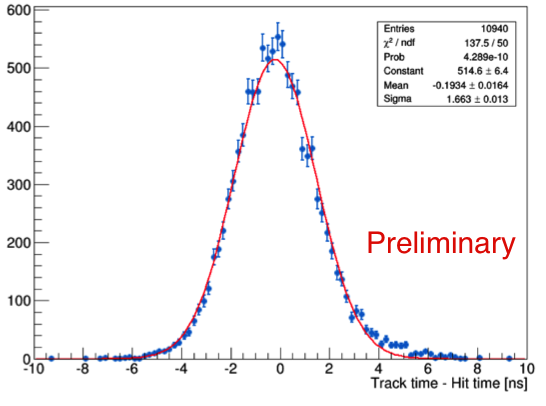}
\caption{Difference between the track time and hit time.}
\label{fig:hit-time}
\end{figure}

\begin{figure}[]
\centering
\includegraphics[width=3.0in]{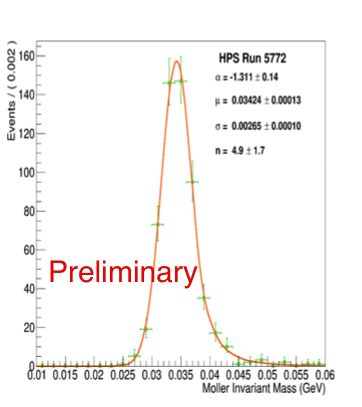}
\caption{Invariant mass of selected e$^-$e$^-$ pairs compatible with Mo\"ller kinematics in a fraction of the data collected.}
\label{fig:moller}
\end{figure}

\begin{figure}[]
\centering
\includegraphics[width=3.0in]{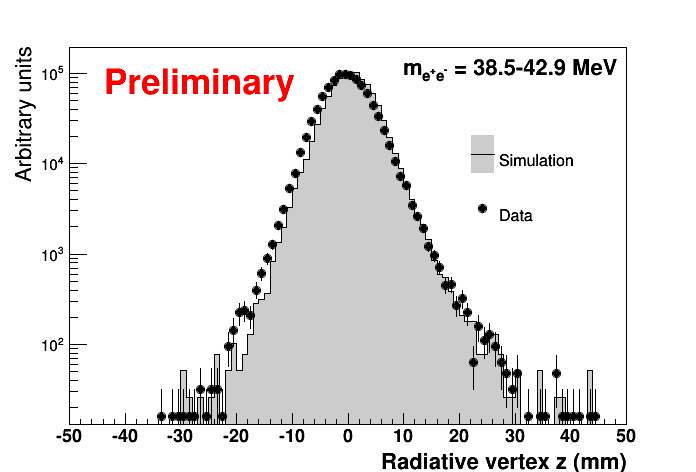}
\caption{Fitted vertex distribution of e$^+$e$^-$ pairs using a QED radiative trident selection for data and simulated events.}
\label{fig:vertex}
\end{figure}

The SVT was operated with 180~V sensor high-voltage bias and had only five bad channels out of 23,004. Beam halo and occupancies were 
measured as expected (see Fig.~\ref{fig:occupancy}). The preliminary results presented below show that key performance observables behave as 
expected at this early stage, where improvements in calibration and alignment are expected. A 2.2~ns time resolution can be extracted from the 
difference between the hit and track time (average of the hit times on the track) seen in Fig.~\ref{fig:hit-time}. 
Further optimizations on the pulse shape fit will likely improve this. For the \Aprime{} searches, the mass and vertex resolutions are key observables. 
The mass resolution is calibrated using e.g. M\"oller events, see Fig.~\ref{fig:moller}; the data show good agreement with the expected distribution 
from simulated events. 
The vertex distribution of e$^+$e$^-$ pairs after radiative QED trident selections in Fig.~\ref{fig:vertex} show reasonable agreement with simulation 
at this stage. Understanding the tails of these distributions is fundamental to the displaced vertex search. 
Results from this run are expected in 2016. 

\subsection{Summary and Outlook}

The SVT for the HPS experiment was designed and built during 2014. It has six layers of silicon microstrip sensor pairs with 50-100~mrad stereo 
angle distributed over approximately 80~cm and sits inside a dipole magnetic field. 
It is placed only 10~cm behind a thin tungsten target and with sensor edges only 0.5~mm from the 
beam to achieve the prompt e$^+$e$^-$ vertex rejection and the acceptance close to the beam. 

The SVT was successfully commissioned together with the whole HPS experiment at nominal operating conditions during a run in the spring of 
2015. Preliminary results show expected occupancies and backgrounds with reasonable agreement between data and simulated events for 
key performance observables such as the mass and vertex resolution. Results are expected in the spring of 2016.

The SVT has been kept cold since the run to mitigate reverse annealing in the irradiated sensors in preparation for continuing running in 2016.

\vfill

%\appendices
%\section{}
%If  I need an appendix it should go here.

\section*{Acknowledgment}
The authors are grateful for the support from Hall~B at JLab and especially the Hall~B engineering 
group for support during installation and decommissioning. They also would like to commend the 
CEBAF personnel for good beam performance during the spring run of 2015. 
The tremendous support from home institutions and supporting staff also needs praise from the 
authors. 
%Work supported by the U.S. Department of Energy under contract number DE-AC02-76SF00515, 
%the National Science Foundation,  
%French Centre National de la Recherche Scientifique and 
%Italian Istituto Nazionale di Fisica Nucleare. 
%Rouven Essig is supported in part by the Department of Energy Early Career research program 
%DESC0008061and by a Sloan Foundation Research Fellowship. Authored by Jefferson Science 
%Associates, LLC under under U.S. Department of Energy contract No. DE-AC05-06OR23177.


%\bibliographystyle{IEEEtran}

\begin{thebibliography}{1}





\bibitem{pamela}
O. Adriani {\it et al.} [PAMELA Collaboration], Nature {\bf 458}, 607 (2009) 
%[arXiv:0810.4995 [astro- ph]],
%\bibitem{pamela2} 
%O. Adriani et al. [PAMELA Collaboration], Phys. Rev. Lett. 106, 201101 (2011) [arXiv:1103.2880 [astro-ph.HE]].
\bibitem{fermi} 
M. Ackermann {\it et al.} [Fermi LAT Collaboration], Phys. Rev. D {\bf 82}, 092004 (2010) 
%[arXiv:1008.3999 [astro-ph.HE]].
%\bibitem{fermi2} 
%M. Ackermann et al. [The Fermi LAT Collaboration], Phys. Rev. Lett. 108, 011103 (2012) [arXiv:1109.0521 [astro-ph.HE]].
%\bibitem{atic} 
%J. Chang et al., Nature 456, 362 (2008).
%\bibitem{hess1} 
%F. Aharonian et al. [H.E.S.S. Collaboration], Phys. Rev. Lett. 101, 261104 (2008) [arXiv:0811.3894 [astro-ph]].
%\bibitem{hess2} 
%F. Aharonian et al. [H.E.S.S. Collaboration], Astron. Astrophys. 508, 561 (2009) [arXiv:0905.0105 [astro-ph.HE]].
\bibitem{nima}
N. Arkani-Hamed, D. P. Finkbeiner, T. R. Slatyer and N. Weiner, Phys. Rev. D {\bf 79}, 015014 (2009).
%M. Pospelov and A. Ritz, 
%ÒAstrophysical Signatures of Secluded Dark Matter,Ó 
%Phys. Lett. B 671 (2009) 391 [arXiv:0810.1502 [hep-ph]],
%M. Cirelli, M. Kadastik, M. Raidal and A. Strumia, Nucl. Phys. B 813, 1 (2009) [arXiv:0809.2409 [hep-ph]].,
%I. Cholis, D. P. Finkbeiner, L. Goodenough and N. Weiner, JCAP 0912, 007 (2009) [arXiv:0810.5344 [astro-ph]], I. Cholis, G. Dobler, D. P. Finkbeiner, L. Goodenough and N. Weiner, Phys. Rev. D 80, 123518 (2009) [arXiv:0811.3641 [astro-ph]].
\bibitem{holdom}
B. Holdom, Phys. Lett. B {\bf 166}, 196 (1986),
%\bibitem{galison}
P. Galison {\it et al,} 
%ÒTwo ZÕs Or Not Two ZÕs?,Ó 
Phys. Lett. B {\bf 136} (1984) 279
\bibitem{Hewett:2012ns} 
  J.~L.~Hewett, {\it et al.},
  ``Fundamental Physics at the Intensity Frontier,''
  arXiv:1205.2671 [hep-ex].
  %%CITATION = ARXIV:1205.2671;%%
%\bibitem{apex1} 
%R.Essig et al, JHEP1102,009(2011)[arXiv:1001.2557 [hep-ph]].
%\bibitem{darkforces}
%ÒDark2012: Dark Forces at AcceleratorsÓ, http://www.lnf.infn.it/conference/dark/index.php
%\bibitem{apex} 
%S. Abrahamyan et al. [APEX Collaboration], Phys. Rev. Lett. 107, 191804 (2011) [arXiv:1108.2750 [hep-ex]].
%\bibitem{mami} 
%H. Merkel et al. [A1 Collaboration], Phys. Rev. Lett. 106, 251802 (2011) [arXiv:1101.4091 [nucl-ex]].
%\bibitem{darklight}  
%M. Freytsis et al, ÒDark Force Detection in Low Energy E-P Colli- sions,Ó JHEP 1001 (2010) 111 [arXiv:0909.2862 [hep-ph]].

\bibitem{proposal_full}
A. Grillo {\it et al.} [HPS Collaboration], JLab PAC37 PR-11-006, http://www.jlab.org/exp prog/PACpage/PAC37/proposals/Proposals/

\bibitem{bible} J. D. Bjorken, R. Essig, P. Schuster and N. Toro, Phys. Rev. D {\bf 80}, 075018 (2009) [arXiv:0906.0580 [hep-ph]].
%\bibitem{hiddensector1} M. Goodsell and A. Ringwald, ÒLight hidden-sector U(1)s in string compactifications,Ó Fortsch. Phys. 58, 716 (2010) [arXiv:1002.1840 [hep-th]].
%\bibitem{hiddensector1}  P. Candelas, G. T. Horowitz, A. Strominger and E. Witten, ÒVacuum Configurations for Superstrings,Ó Nucl. Phys. B 258, 46 (1985).
%\bibitem{hiddensector1}  E. Witten, ÒNew Issues in Manifolds of SU(3) Holonomy,Ó Nucl. Phys. B 268, 79 (1986).
%\bibitem{hiddensector1} S. Andreas, M. D. Goodsell and A. Ringwald, ÒDark matter and Dark Forces from a super-symmetric hidden sector,Ó arXiv:1109.2869 [hep-ph].
%\bibitem{hiddensector1} J. Jaeckel and A. Ringwald, ÒThe Low-Energy Frontier of Particle Physics,Ó Ann. Rev. Nucl. Part. Sci. 60, 405 (2010) [arXiv:1002.0329 [hep-ph]].
%\bibitem{mass1} P. Fayet, ÒU-Boson Production in E+ E- Annihilations, Psi and Upsilon Decays, and Light Dark Matter,Ó Phys. Rev. D 75 (2007) 115017 [arXiv:hep-ph/0702176].
%\bibitem{mass1} C. Cheung, J. T. Ruderman, L. T. Wang and I. Yavin, ÒKinetic Mixing as the Origin of Light Dark Scales,Ó Phys. Rev. D 80 (2009) 035008 [arXiv:0902.3246 [hep-ph]].
%\bibitem{mass1} N. Arkani-Hamed and N. Weiner, JHEP 0812, 104 (2008) [arXiv:0810.0714 [hep-ph]].
%\bibitem{mass1} D. E. Morrissey, D. Poland and K. M. Zurek, ÒAbelian Hidden Sectors at a Gev,Ó JHEP 0907 (2009) 050 [arXiv:0904.2567 [hep-ph]].
%\bibitem{g-2_constraints}
%M.~Pospelov, Phys.\ Rev.\ D {\bf 80}, 095002 (2009) [arXiv:0811.1030];
 %G. W. Bennett et al., [Muon G2 Collaboration], Phys. Rev. D{\bf 73} 072003 (2006)[hep?ex/0602035].
%\bibitem{Endo:2012hp} M.~Endo, K.~Hamaguchi and G.~Mishima, Phys.\ Rev.\ D {\bf 86}, 095029 (2012) [arXiv:1209.2558 [hep-ph]].
%\bibitem{cmb} J. B. Dent, F. Ferrer and L. M. Krauss, 
%ÒConstraints on Light Hidden Sector Gauge Bosons from Supernova Cooling,Ó arXiv:1201.2683 [astro-ph.CO].
%\bibitem{fixedtargetexp} J. D. Bjorken et al., Phys. Rev. D 38 (1988) 3375, E. M. Riordan et al., Phys. Rev. Lett. 59 (1987) 755, A. Bross, M. Crisler, S. H. Pordes, J. Volk, S. Errede and J. Wrbanek, Phys. Rev. Lett. 67 (1991) 2942.
%\bibitem{collider}
%KLOE-2 Collaboration, %ÒSearch for a Vector Gauge Boson in Phi Meson Decays with theKLOE Detector,Ó Phys. Lett. B 706 (2012) 251 [arXiv:1110.0411 [hep-ex]], 
%M. Reece and L. T. Wang, 
%ÒSearching for the Light Dark Gauge Boson in Gev-Scale Experi-ments,Ó JHEP 0907 (2009) 051 [arXiv:0904.1743 [hep-ph]],
%B. Aubert et al. [BABAR Collaboration], Phys. Rev. Lett. 103, 081803 (2009) [arXiv:0905.4539 [hep-ex]].
\bibitem{fadc250}
E.Jastrzembski and H. Dong, https://coda.jlab.org/drupal/content/vme-payload-modules
%https://coda.jlab.org/drupal/system/files/pdfs/HardwareManual/fADC250/FADC250%20data%20format%20V23.pdf
\bibitem{dose}
Rashevskaya {\it et~al.}, Radiation damage of silicon structures with electrons of 900-MeV, Nucl. Instr. Meth. A, {\bf 485}, 126-132, (2002)
\bibitem{d0run2b}
D. S. Denisov and S. Soldner-Rembold, FERMILAB-PROPOSAL-0925.
\bibitem{apv25}
M.J. French {\it et al.}, Design and results from the APV25, a deep sub-micron CMOS front-end chip for the CMS tracker , Nucl. Instr. Meth. A, {\bf 466}, 359-365 (2001)
\bibitem{Friedl:2009zz}
Friedl {\it et~al.} (2009), Readout of silicon strip detectors with position and timing information, Nucl. Instr. Meth. A, {\bf 598}, 82-83 (2009)
\bibitem{paper_testrun}
P. Hansson Adrian {\it et al.} [HPS Collaboration], The Heavy Photon Search test detector, Nucl. Instr. Meth. A 777 (2015), 91-101, http://dx.doi.org/10.1016/j.nima.2014.12.017
\bibitem{coda}
https://coda.jlab.org
%M. Battaglieri, S. Boyarinov, S. Bueltmann, V. Burkert, A. Celentano, G. Charles, W. Cooper, C. Cuevas, N. Dashyan, R. DeVita, C. Desnault, A. Deur, H. Egiyan, L. Elouadrhiri, R. Essig, V. Fadeyev, C. Field, A. Freyberger, Y. Gershtein, N. Gevorgyan, F.-X. Girod, N. Graf, M. Graham, K. Griffioen, A. Grillo, M. Guidal, G. Haller, P. Hansson Adrian, R. Herbst, M. Holtrop, J. Jaros, S. Kaneta, M. Khandaker, A. Kubarovsky, V. Kubarovsky, T. Maruyama, J. McCormick, K. Moffeit, O. Moreno, H. Neal, T. Nelson, S. Niccolai, A. Odian, M. Oriunno, R. Paremuzyan, R. Partridge, S.K. Phillips, E. Rauly, B. Raydo, J. Reichert, E. Rindel, P. Rosier, C. Salgado, P. Schuster, Y. Sharabian, D. Sokhan, S. Stepanyan, N. Toro, S. Uemura, M. Ungaro, H. Voskanyan, D. Walz, L.B. Weinstein, B. Wojtsekhowski, The Heavy Photon Search test detector, Nuclear Instruments and Methods in Physics Research Section A: Accelerators, Spectrometers, Detectors and Associated Equipment, Volume 777, 21 March 2015, Pages 91-101, ISSN 0168-9002, http://dx.doi.org/10.1016/j.nima.2014.12.017

%\bibitem{proposal_testrun}
%P. Hansson Adrian {\it et al.} [HPS Collaboration], HPS Test Run Proposal and PAC39 Update, https://confluence.slac.stanford.edu/display/hpsg/Project+Overview,
%\bibitem{pac39}
%P. Hansson et al. [HPS Collaboration], HPS Update PAC 39, https://confluence.slac.stanford.edu/display/hpsg/Project+Overview
%\bibitem{egs5}
%H. Hirayama, Y. Namito, A.F. Bielajew, S.J. Wilderman and W.R. Nelson, SLAC-R-730 (2005) and KEK Report 2005-8 (2005).


\end{thebibliography}
%\begin{thebibliography}{1}
%\bibitem{IEEEPDFRequirement401}
%IEEE Content Engineering, \emph{PDF Specification for IEEE Xplore}. Available: http://www.ieee.org/portal/cms\_docs/pubs/confstandards/pdfs/IEEE-PDF-SpecV401.pdf.
%
%\end{thebibliography}

\end{document}